\documentclass[prl,preprint,onecolumn,showpacs,floatfix,superscriptaddress]{revtex4-2}

\usepackage{graphicx}
\usepackage{subfigure}
\usepackage{bm}
\usepackage{amsmath}
\usepackage{tabularx}
\usepackage{array}
\usepackage{multirow}
\usepackage{float}
\usepackage{xcolor}
\usepackage{hyperref}
\usepackage{physics}
\usepackage{lineno}

\begin{document}


\title{Room-temperature waveguide integrated quantum register in a semiconductor photonic platform}
\author{Haibo Hu}
\thanks{These authors contributed equally.}
\affiliation{Ministry of Industry and Information Technology Key Lab of Micro-Nano Optoelectronic Information System, Guangdong Provincial Key Laboratory of Semiconductor Optoelectronic Materials and Intelligent Photonic Systems, Harbin Institute of Technology, Shenzhen, 518055, China}
\affiliation{Pengcheng Laboratory, Shenzhen 518055, China}

\author{Yu Zhou\textsuperscript{$\dagger$}}
\thanks{These authors contributed equally.}
\affiliation{Ministry of Industry and Information Technology Key Lab of Micro-Nano Optoelectronic Information System, Guangdong Provincial Key Laboratory of Semiconductor Optoelectronic Materials and Intelligent Photonic Systems, Harbin Institute of Technology, Shenzhen, 518055, China}
\affiliation{Quantum Science Center of Guangdong-HongKong-Macao Greater Bay Area (Guangdong), Shenzhen 518045, China}
\author{Ailun Yi}
\thanks{These authors contributed equally.}
\affiliation{National Key Laboratory of Materials for Integrated Circuits, Shanghai Institute of Microsystem and Information Technology, Chinese Academy of Sciences, Shanghai 200050, China}
\affiliation{The Center of Materials Science and Optoelectronics
Engineering, University of Chinese Academy of Sciences, Beijing 100049, China.}
\author{Tongyuan Bao}
\affiliation{Ministry of Industry and Information Technology Key Lab of Micro-Nano Optoelectronic Information System, Guangdong Provincial Key Laboratory of Semiconductor Optoelectronic Materials and Intelligent Photonic Systems, Harbin Institute of Technology, Shenzhen, 518055, China}
\author{Chengying Liu}
\affiliation{Ministry of Industry and Information Technology Key Lab of Micro-Nano Optoelectronic Information System, Guangdong Provincial Key Laboratory of Semiconductor Optoelectronic Materials and Intelligent Photonic Systems, Harbin Institute of Technology, Shenzhen, 518055, China}
\author{Qi Luo}
\affiliation{Ministry of Industry and Information Technology Key Lab of Micro-Nano Optoelectronic Information System, Guangdong Provincial Key Laboratory of Semiconductor Optoelectronic Materials and Intelligent Photonic Systems, Harbin Institute of Technology, Shenzhen, 518055, China}
\author{Yao Zhang}
\affiliation{Ministry of Industry and Information Technology Key Lab of Micro-Nano Optoelectronic Information System, Guangdong Provincial Key Laboratory of Semiconductor Optoelectronic Materials and Intelligent Photonic Systems, Harbin Institute of Technology, Shenzhen, 518055, China}
\author{Zi Wang}
\affiliation{Ministry of Industry and Information Technology Key Lab of Micro-Nano Optoelectronic Information System, Guangdong Provincial Key Laboratory of Semiconductor Optoelectronic Materials and Intelligent Photonic Systems, Harbin Institute of Technology, Shenzhen, 518055, China}
\author{Qiang Li}
\affiliation{Institute of Advanced Semiconductors, Zhejiang Provincial Key Laboratory of Power Semiconductor Materials and Devices, ZJU-Hangzhou Global Scientific and Technological Innovation Center, Hangzhou, Zhejiang 311200, China}
\affiliation{State Key Laboratory of Silicon Materials and Advanced Semiconductors $\&$ School of Materials Science and Engineering, Zhejiang University, Hangzhou 310027, China}
\author{Dawei Lu}
\affiliation{Quantum Science Center of Guangdong-HongKong-Macao Greater Bay Area (Guangdong), Shenzhen 518045, China}
\affiliation{Shenzhen Institute for Quantum Science and Engineering and Department of Physics, Southern University of Science and Technology, Shenzhen 518055, China}
\author{Zhengtong Liu}
\affiliation{Pengcheng Laboratory, Shenzhen 518055, China}
\author{Shumin Xiao}
\affiliation{Ministry of Industry and Information Technology Key Lab of Micro-Nano Optoelectronic Information System, Guangdong Provincial Key Laboratory of Semiconductor Optoelectronic Materials and Intelligent Photonic Systems, Harbin Institute of Technology, Shenzhen, 518055, China}
\affiliation{Pengcheng Laboratory, Shenzhen 518055, China}
\affiliation{Quantum Science Center of Guangdong-HongKong-Macao Greater Bay Area (Guangdong), Shenzhen 518045, China}
\affiliation{Collaborative Innovation Center of Extreme Optics, Shanxi University, Taiyuan, Shanxi 030006, China}
\author{Xin Ou\textsuperscript{$\dagger$}}

\affiliation{National Key Laboratory of Materials for Integrated Circuits, Shanghai Institute of Microsystem and Information Technology, Chinese Academy of Sciences, Shanghai 200050, China}
\affiliation{The Center of Materials Science and Optoelectronics
Engineering, University of Chinese Academy of Sciences, Beijing 100049, China.}
\author{Qinghai Song}
\thanks{Corresponding authors: Yu Zhou(zhouyu2022@hit.edu.cn) or Xin Ou (ouxin@mail.sim.ac.cn) or Qinghai Song 
 (qinghai.song@hit.edu.cn)}

\affiliation{Ministry of Industry and Information Technology Key Lab of Micro-Nano Optoelectronic Information System, Guangdong Provincial Key Laboratory of Semiconductor Optoelectronic Materials and Intelligent Photonic Systems, Harbin Institute of Technology, Shenzhen, 518055, China}
\affiliation{Pengcheng Laboratory, Shenzhen 518055, China}
\affiliation{Quantum Science Center of Guangdong-HongKong-Macao Greater Bay Area (Guangdong), Shenzhen 518045, China}
\affiliation{Collaborative Innovation Center of Extreme Optics, Shanxi University, Taiyuan, Shanxi 030006, China}

\begin{abstract}

{\textbf{Quantum photonic integrated circuits are reshaping quantum networks and sensing by providing compact, efficient platforms for practical quantum applications. Despite continuous breakthroughs, integrating entangled registers into photonic devices on a CMOS-compatible platform presents significant challenges. Herein, we present single electron-nuclear spin entanglement and its integration into a silicon-carbide-on-insulator (SiCOI) waveguide. We demonstrate the successful generation of single divacancy electron spins and near-unity spin initialization of single $^{13}$C nuclear spins. Both single nuclear and electron spin can be coherently controlled and a maximally entangled state with a fidelity of 0.89 has been prepared under ambient conditions. Based on the nanoscale positioning techniques, the entangled quantum register has been further integrated into SiC photonic waveguides for the first time. We find that the intrinsic optical and spin characteristics of the register are well preserved and the fidelity of the entangled state remains as high as 0.88. Our findings highlight the promising prospects of the SiCOI platform as a compelling candidate for future scalable quantum photonic applications.}
}

\end{abstract}

\flushbottom
\maketitle
\thispagestyle{empty}
\pagebreak[3]

~\\
~\\
~\\
\section*{Introduction}

Nuclear-electron quantum registers are essential components of multi-node quantum networks\cite{pompili2021realization,stas2022robust} and quantum sensing\cite{jarmola2021demonstration,soshenko2021nuclear,simin2016all}. These registers, with electron-coupled nuclear spins serving as memory qubits or sensors, play a crucial role in network operations\cite{pompili2021realization,stas2022robust,simin2016all}. Color centers in diamonds have been at the forefront of constructing quantum networks, with significant advancements and achievements realized over the past decades with two main approaches \cite{bernien2013heralded,hensen2015loophole,hermans2022qubit,pompili2021realization,sipahigil2016integrated,stas2022robust,bersin2023development}. The first approach entails the entanglement of nitrogen-vacancy (NV) centers in bulk diamonds at distant sites using herald protocols\cite{bernien2013heralded,hensen2015loophole,hermans2022qubit,pompili2021realization}. A more integrated strategy incorporates silicon vacancy\cite{sipahigil2016integrated}  and proximal nuclear spins\cite{stas2022robust} within diamond nanophotonics, marking significant steps in quantum network construction. At the same time, quantum sensing using defects in diamonds has advanced significantly, finding applications across diverse areas and demonstrating considerable development\cite{shi2015single}. To date, defect-based quantum systems are rapidly emerging as one of the cornerstones for the next generation of integrated quantum networking and sensing. Despite significant advancements in these quantum technologies, achieving a fully integrated quantum photonic network chip with substantial volume and energy advantages remains a formidable challenge. The primary difficulty lies in integrating entangled registers into photonic devices on a CMOS-compatible platform, a task that continues to pose substantial obstacles.

Silicon carbide, an emerging CMOS-compatible quantum material\cite{falk2013polytype,widmann2015coherent,babin2022fabrication}, holds the potential to be the host for realizing a fully monolithic quantum photonic network processor\cite{lukin2020integrated,awschalom2021development}. This potential is attributed to the recent rapid advancements in SiC quantum registers and SiCOI platforms. On the one hand, the coherence time of individual electron spins in SiC has significantly extended from milliseconds\cite{christle2015isolated} to seconds\cite{anderson2022five}. Besides the electron spin, wealthy nuclear spin resources are ideal quantum memory candidates in quantum network construction\cite{hermans2022qubit}. Ensemble nuclear spins in SiC have been efficiently initialized through dynamical nuclear polarization (DNP)\cite{falk2015optical}. Furthermore, coherent control and entanglement of both ensemble\cite{klimov2015quantum} and individual nuclear spins\cite{bourassa2020entanglement} at cryogenic temperatures provides a solid foundation for many nuclear-spin involving quantum networks and sensing applications\cite{taminiau2014universal,ajoy2015atomic, jarmola2021demonstration,soshenko2021nuclear,arrad2014increasing,hirose2016coherent}. On the other hand, the advancement of quantum-grade SiCOI platforms has broadened opportunities for integrating photonic devices directly on-chip. SiCOI platforms are recognized for their versatile functionalities, attributed to the inherent nonlinear properties of SiC, including photon frequency conversion to the telecom band\cite{lukin20204h}, generation of entangled photon pairs\cite{guidry2022quantum}, and electro-optic modulation\cite{powell2022integrated}. Despite continuous breakthroughs in SiC quantum registers and the SiCOI photonic platform, integrating them into a quantum photonics platform with waveguide-integrated entangled quantum registers remains a significant challenge, leaving this area still blank. Here, we address this challenge by experimentally demonstrating the integration of single electron-nuclear spin entanglement into a silicon-carbide-on-insulator (SiCOI) waveguide. Our findings indicate a promising approach for achieving scalable quantum photonic applications in the future.
\\

\section*{Results}
\noindent\textbf{Generation and coherent control of single divacancy spins on SiCOI}

PL6 is a relatively new type of 4H-SiC divacancy color center with attractive optical and spin properties like high brightness (150 kcps) and ODMR contrast (30$\%$) at room temperature\cite{li2022room}. PL6 is believed to be situated within stacking faults acting as quantum wells\cite{ivady2019stabilization}. This unique positioning enhances these defects'(including PL5 and PL7) stability against photo-ionization, different from the previously identified $V_{Si}V_{C}$ defects (PL1-PL4)\cite{falk2013polytype}. To maintain the key characteristics of the single divacancy color centers, the SiCOI wafer is fabricated by a thinning and polishing technique\cite{lukin20204h}. First, a 4H-SiC wafer with an epitaxy layer is bonded to an oxidized Si wafer. Subsequently, the bonded SiC layer is mechanically ground and then subjected to chemical-mechanical polished (CMP) to achieve a thickness of several micrometers. The SiCOI wafer with a designed thickness of 200 nm is obtained after etching the SiC layer(see Methods and Supplementary Fig. 1 for more details of the sample). The divacancy spins are generated in SiCOI via a selective carbon ion implantation, as displayed in Fig. 1a. 8$\times$8 hole arrays are fabricated on the SiC layer using 180 nm thick PMMA (A4) as a photoresist through E-beam lithography (Fig. 1a). Following the 30 keV Carbon ion implantation with a dose of 7.8$\times$10$^{11}$cm$^{-2}$, the sample is annealed in a high vacuum at 900 °C for 30 minutes to remove residual lattice damage. To characterize the defects array, a single-defect direct imaging system (Supplementary Fig. 3a) is built, and the acquired image taken by an InGaAs camera is shown in Fig. 1b. The designed defect array is clear with the same spacing length and numbers as the designed pattern (Supplementary Fig. 3b). Confocal photoluminescence scanning of the same area is then performed using off-resonant excitation at continuous wave (CW) 914 nm (Supplementary Note 1 for the setups). The two maps in Fig. 1b acquired by different imaging methods are similar, reaffirming that we have successfully generated the defect array in SiCOI as designed. 

To confirm the defect type, optically detected magnetic resonance (ODMR) of the defect is performed at different B fields parallel to the c-axis. Fig. 1c summarizes the results of the defect marked with a red circle in Fig. 1b. The two branches of the ODMR spectrum diverge with a slope of 2.82 $\pm$ 0.02 MHz per gauss, the uncertainty is from the fitting of the peak positions in Fig.1c. The single photon emission nature is further confirmed by the second-order correlation measurement presented in Supplementary Fig. 4a.  Together with the zero-field splitting parameters D = 1340.4 $\pm$ 1.3 MHz, E = 6.95 $\pm$ 1.82 MHz, 168.8 $\pm$ 2.5 Kcps saturation intensity (Supplementary Fig. 4, the uncertainties arise from the fitting process), all these characteristics are consistent with PL6\cite{li2022room}. The small discrepancies between the D and E with the literature are attributed to the lateral and longitudinal strain of the thin film\cite{falk2014electrically}. Subsequently, coherent control of the single PL6 electron spin has been demonstrated, as evidenced by the Rabi oscillation and Ramsey fringes displayed in Figs. 1d and 1e, respectively. From the fitting, the inhomogeneous spin-dephasing time T$^{*}_{2}$ is extracted to be $460 \pm 35 $ ns. The longitudinal relaxation time T$_1$ is measured to be 146 $\pm$ 44 $\mu$s (Supplementary Fig. 4), similar to those in bulk materials\cite{li2022room}. The coherence time T$_2$ is measured to be 4.62 $\pm$ 1.65 $\mu$s (Supplementary Fig. 4), lower than 23.2 $\mu$s in the bulk material\cite{li2022room}. The coherence time is influenced by various factors like isotopic composition in the solid, surface treatment methodologies, and the ion implantation processes. The sample we used in this work is a primitive commercial SiC wafer. The coherence time can be effectively extended through the application of dynamical decoupling techniques\cite{de2010universal} and isotopic engineering \cite{bourassa2020entanglement}. 

\noindent\textbf{Coherent control of single nuclear spins on SiCOI}

Nuclear spins are crucial resources in defect-based quantum technologies. Based on the controlled generation and coherent control of a single PL6 spin, we proceed to nuclear spin manipulation and initialization. In natural SiC, approximately 1.1$\%$ of the carbon atoms ($^{13}$C) and 4.7$\%$ of silicon atoms ($^{29}$Si) possess nuclear spins with I = 1/2. Strong coupling occurs when the nuclear spin resides within several lattice sites of PL6. The hyperfine interaction can exceed both the electron and nuclear spin dephasing rates 1/T$^{*}_{2}$. This strong coupling is valuable for implementing fast gate operations and high-speed quantum memories\cite{rao2016characterization}. We successfully identified an electron-nuclear spin-coupled system PL6 B in the sample, as shown in Fig. 2a. The single PL6 electron spin is strongly coupled to a first shell $^{13}$C nuclear spin\cite{rao2016characterization}. We use 0, ±1, and $\uparrow$, $\downarrow$ to denote the electron and nuclear spin, respectively. The full Hamiltonian of the system can be written as 
\begin{equation}
\label{eq:1}
H= D \cdot [S_z^2 - S(S + 1)/3] + {\gamma _e}\mathbf{B} \cdot \mathbf{S} + \mathbf{S} \cdot \mathbf{A} \cdot \mathbf{I} - {\gamma _{{}^{13}C}}\mathbf{B} \cdot \mathbf{I}
\end{equation}
Here, D represents the zero-field splitting parameter of the electron spin, $\rm{S}$ and $\rm{I}$ are the electron spin-1 and nuclear spin-1/2 operators, respectively. $\gamma _e =2.8\rm{\thinspace GHz \thinspace T^{-1}}$ is the electron-spin gyromagnetic ratio. ${\gamma_{{}^{13}C}=10.708\rm{\thinspace  MHz\thinspace T^{-1}}}$ is the gyromagnetic ratio of $^{13}$C nuclear spin. $\rm{A}$ is the hyperfine-interaction tensor between PL6 and $^{13}$C. When the nuclear spin resides in the first shell, the electron-nuclear spin coupling is strong enough to split the ground state energy levels. Two distinct, individually addressable transitions in SiC, separated by approximately 55.1 $\pm$ 0.6 MHz are identified. These transitions correspond to the ${\uparrow}$ and  ${\downarrow}$ nuclear spin states, observed without an external magnetic field. The hyperfine interaction strength varies drastically depending on their position in the lattice. We attribute this huge hyperfine splitting to coupling with a carbon isotope nuclear spin $^{13}$C in the first shell\cite{son2006divacancy}. When a c-axis magnetic field is added, m$_s\pm1$ are no longer degenerate and split at a slope of 2.82 $\rm{MHz/G}$ due to Zeeman splitting, as shown in Figs. 2b and 2c. The ground-state spin-level anticrossings (GSLAC) region is marked with a dashed circle. 

We chose a B field at around 200 G between the GSLAC point and zero fields to resolve the four transitions well for the following coherent control. The simplified representation of the energy levels of the electron-nuclear coupled system at B = 200 G is depicted in Fig. 2d. Since it is far away from the GSLAC point, ${\ket{0,\uparrow}}$ and ${\ket{0,\downarrow}}$ are degenerate.  There is no nuclear spin polarization at this B field, so the nuclear spin must be mapped to the electron spin before control and readout, as the control sequence depicted in Figs. 2e and 2f. The frequency of RF corresponds to the energy gap between ${\ket{-1,\uparrow}}$ and ${\ket{-1,\downarrow}}$. The electron spin is first prepared in the $m_{s} = -1$ state, and an RF pulse is used to drive nuclear Rabi oscillations (Fig. 2e) and Ramsey fringes (Fig. 2f). The Ramsey fringes yield a pure dephasing time of 13 $\pm$ 4 $\mu$s. 
The observed pure dephasing time $T_{2}^{*}$ of 13 $\mu$s is notably shorter than the millisecond range\cite{maurer2012room}, and the sustained Rabi oscillations lasting up to 8 $\mu$s without significant decay are also brief compared to the hundreds of microseconds \cite{babin2022fabrication,bourassa2020entanglement}. This relatively short timescale can be attributed to the strong interaction with the environment through the electron spin\cite{mizuochi2009coherence}. It can, however, be significantly extended by selecting a nuclear spin not in the first shell or by decoupling the electron spin via the ionization process\cite{dutt2007quantum}. Furthermore, the SiC wafer used in our study is a basic commercial sample. Coherence times for both nuclear and electron spins can be effectively enhanced through dynamical decoupling techniques and isotopic engineering\cite{bourassa2020entanglement}. Besides a single $^{13}$C nuclear spin, a single $^{29}$Si nuclear spin has also been coherently controlled on SiCOI, as presented in Supplementary Note 3.
~\\

\noindent\textbf{Near-unity Dynamic Polarization of a Single Nuclear Spin on SiCOI}

In addition to coherent control, efficient initialization of the single nuclear spin memory is another critical capability in quantum information processing. Dynamical nuclear polarization (DNP) can effectively transfer electron spins to neighbouring nuclei via hyperfine interaction\cite{falk2015optical,jacques2009dynamic}, serving as a basis for quantum memories and computing\cite{dutt2007quantum,cai2021parallel}, nuclear magnetic resonance sensitivity enhancement\cite{giovannetti2011advances}, and solid-state nuclear gyroscopes\cite{jarmola2021demonstration}. We demonstrate that room-temperature DNP can be efficiently utilized on SiCOI, achieving a near-unity degree of polarization for a single nuclear spin. When the system is far from the anticrossing point, nuclear spins will not be optically polarized, and equal populations of ${\ket{\pm 1,\uparrow}}$ and ${\ket{\pm 1,\downarrow}}$ will result in the same intensity of the ODMR spectrum, as shown in Fig. 2c. Near GSLAC, in each optical cycle, ${\ket{0,\downarrow}}$ may evolve into ${\ket{-1,\uparrow}}$, exchanging electron and nuclear polarizations. Subsequent optical cycles polarize ${\ket{-1,\uparrow}}$ to ${\ket{0,\uparrow}}$. The whole process polarizes the nuclear spins simply by optical illumination\cite{falk2015optical,jacques2009dynamic}.

To uncover the detailed mechanism of the DNP near the GSLAC point , ODMR measurements at different magnetic fields have been conducted with a 914 nm off-resonant CW excitation as illustrated in Fig. 3a. Our microwave source limits the lower bound of the microwave frequency to 20 MHz. Once the hyperfine tensor A is given, the energy separation between the electron and nuclear eigenstates can be calculated from the total Hamiltonian\cite{wang2013sensitive,ivady2015theoretical}. The most fitted results are presented with colored solid lines (with $A_{xx}=A_{yy}=93.1 \pm 0.3 $ MHz, $A_{zz}=56.5 \pm 0.2$ MHz), where each color line represents a transition. Notably, ${\ket{-1,\uparrow}}$ to ${\ket{0,\uparrow}}$ maintains its intensity as the magnetic field approaches the anticrossing point. In contrast, ${\ket{-1,\downarrow}}$ to ${\ket{0,\uparrow}}$ weakens and vanishes from B = 460 G, signifying where strong nuclear polarization occurs. The degree of nuclear spin polarization P is defined as $({I^ + } - {I^ - })/({I^ + } + {I^ - })$, where $I^+$ and $I^-$ denote the populations of the nuclear spins $\uparrow$ and $\downarrow$ respectively\cite{falk2015optical,jacques2009dynamic}. The populations are quantified by performing the Lorentzian fit of individual peaks. As illustrated in Fig.3b, the solid lines are the theoretically calculated curve of $\rho={\rho }_{\uparrow}-{\rho }_{\downarrow}$ ($\rho{\uparrow}$ and $\rho{\downarrow}$ are the populations of nuclear spin up and down, see Methods). The maximum polarization degree achieved is $0.98\pm 0.04$, reaching near-unity initialization of the single nuclear spin. To confirm this polarization, the ODMR spectra and Rabi oscillations near and far away from the GSLAC point are compared as represented in Figs. 3c-e. When B = 217 G, the two ODMR spectra have similar contrast (Fig. 3c), the same as in Rabi oscillations(Fig. 3d). However, when B = 478 G where the nuclear spin polarization is strong, ${\ket{-1,\downarrow}}$ to ${\ket{0,\uparrow}}$ vanishes in both ODMR spectra and corresponding Rabi oscillation(Figs. 3e and 3f). 
\\

\noindent\textbf{Optically detected nuclear magnetic resonance (ODNMR) and electron-nuclear entanglement generation}

After performing complete control and near-unity polarization of a single nuclear spin, we have all the prerequisites to demonstrate the generation of an electron-nuclear entangled state of the PL6 B. The simplified energy level scheme of the electron-nuclear coupled system is depicted in Fig. 4a. MW (${\ket{0,\uparrow}}$ to ${\ket{-1,\uparrow}}$) and RF2 (${\ket{0,\uparrow}}$ to ${\ket{0,\downarrow}}$) in this subspace can be obtained directly from the ODMR measurements, while RF1 (${\ket{-1,\uparrow}}$ to ${\ket{-1,\downarrow}}$) is missing due to the strong nuclear polarization near the GSLAC point, as also evident in Fig. 3a. To recover RF1, the pulse sequence in Fig. 4b is used after the system is optically initialized to ${\ket{-1,\uparrow}}$, The electron spin is prepared in the $m_s$ = -1 state and then an RF pulse is applied. The state is then read out by projecting onto the electron spin again. By varying the frequency of the RF pulse, The optically detected nuclear magnetic resonance (ODNMR) spectra are obtained as indicated in Fig. 4b. RF1 transition is extracted to be 27.688 $\pm$ 0.009 MHz from the Lorentz fitting. After successfully addressing all three transitions, quantum state tomography (QST) can be conducted utilizing the MW as a working transition(see Supplementary Fig. 6, Supplementary Table 1 and 2 for the details). The evolution of electron-nuclear states is tracked by performing quantum state tomography at different stages. First, the qubits are initialized by optical pumping due to nuclear spin pumping, after which quantum state tomography is performed. The measured density matrix is presented below the circuit in Fig. 4b with a fidelity of 0.98. This fidelity is the same as the degree of polarization obtained with the ODMR peak fitting method in the previous section, thus reconfirming the near-unity nuclear spin polarization. Next, a maximal superposition is created by applying two gates, as presented in Fig. 4d, entangling the electron and nuclear spin into one of the Bell states ${\ket{\Psi^+} = (\ket{0,\uparrow} + \ket{-1,\downarrow}})/\sqrt{2}$ (Fig. 4d). We successfully generated the electron-nuclear entanglement in this system with a fidelity of 0.89, estimated from the density matrix obtained through QST (see Methods and Supplementary Note Tables for details). Similar to PL6 B, PL6 C is another single electron spin with a strongly coupled first-shell $^{13}$C nuclear spin. The coupling strength is nearly 56 MHz as shown in Supplementary Fig. 10a. Both ODNMR and entanglement generation of this qunatum register have been successfully demonstrated within the membrane, as illustrated in Supplementary Fig. 11.
\\
\\
\\
\textbf{Deterministic integration of the entangled quantum register into a waveguide}

A critical step for developing SiCOI quantum photonics platforms is integrating single divacancy spins or quantum registers into photonic circuits and realizing on-chip integrated quantum photonic devices. We have achieved this goal by further developing a nanoscale positioning technique \cite{liu2021nanoscale} via direct camera imaging and pattern recognition (Supplementary Fig. 3). With this technique, the positioning accuracy of a single quantum register has been honed close to 10 nm, enabling the precise design and realization of surrounding photonic structures for on-chip integration. To illustrate this potential, We have designed waveguides equipped with directional couplers that encase the PL6 C quantum register. The photonic nanostructures are then fabricated with a combined process of electron beam lithography and reactive ion etching (Methods). The inset in Fig. 5a shows the top-view scanning electron microscope (SEM) image of one waveguide, and all the structural parameters match the design very well. 

The quantum register was optically characterized by scanning the waveguide confocally. A bright emission spot can be clearly seen in Fig. 5b. Since the surrounding SiC has been completely etched away, we can confirm that the PL6 C quantum register is encased in the waveguide. By fixing the excitation beam on PL6 C, we changed the collection angle with a scanning mirror and recorded the angle-dependent emission, as depicted in Fig. 5c. Two additional emission spots can be observed at the positions of the grating couplers, indicating that the quantum register is integrated inside the waveguide and that its emission is being confined and guided by the nanostructures. Similarly, the spin properties of the quantum register in the photonic waveguide were examined. The ODMR spectrum under different magnetic B fields shows an apparent 56.1 $\pm$ 0.2 MHz splitting (Fig. 5d), confirming the persistent coupling between the nuclear and electron spins.

Based on the successful integration, we compared the stability of photon emission before and after etching. The results are summarized in Fig. 5e as purple hollow rhombus (before integration) and cross (after integration). The photon emission was collected at intervals of 0.1 s over a duration of 300 s. It is easy to see that the emissions both before and after integration are stable, and no blinking or bleaching can be observed. Meanwhile, two characteristic Rabi oscillations (MW transition) are shown in the same figure with green hollow rhombuses (before integration) and crosses (after integration), thus demonstrating that the register can be coherently controlled without degradation. The complementary experiments on spin properties (Supplementary Fig. 10) reveal that the T$_1$ and T$_2^*$ of the waveguide-integrated PL6 C are 188 $\pm$ 43 $\mu$s and 0.94 $\pm$ 0.17 $\mu$s, respectively, comparable to the values before the integration. All these results confirm that the quantum register has been coupled to the waveguide without compromising its spin or optical properties. We attribute this to the size of the waveguide's width, which is usually on the wavelength or subwavelength scale and far larger than the influence of surface noise sources (\textless 100 nm)\cite{sangtawesin2019origins}.

With the established stability, we repeated the near-unity nuclear spin polarization and ODNMR (Supplementary Note 5) as conducted with PL6 B in the membrane before. Near the GSLAC point, we observed the same near-unity polarization as displayed in Supplementary Fig. 10b, achieving a maximum polarization degree of $0.99\pm 0.01$ from the peak fitting. Finally, a Bell state ${\ket{\Psi^+} = (\ket{0,\uparrow} + \ket{-1,\downarrow}})/\sqrt{2}$ has been prepared. The corresponding density matrix is depicted in Fig. 5f, and the fidelity estimated from quantum state tomography is still as high as 0.88, which is similar to the value before integration. The raw data of the quantum state tomography are presented in Supplementary Table 3. Then, we confirm that the quantum register can be integrated into photonic structures without obvious degradation.

In addition to PL6 C, the single electron spin PL6 A has also been integrated into a waveguide with the same technique, as depicted in Supplementary Note 4. No discernible degradation in optical or spin properties has been observed after integration, underscoring the reliability of our approach. Our presented techniques are applicable to other photonic devices as well, e.g., beam splitters\cite{wang2022waveguide}, couplers, and micro rings. As a consequence, more complex quantum register-integrated on-chip quantum photonic circuits can be expected in the near future. Additionally, all single spins can be imaged in real-time using a commercial InGaAs CCD (Supplementary Video 1), enabling potential post-fabrication control of each quantum register in large-scale integration in the future.

~\\

\section*{Discussion}
\hspace{1.0 em} SiCOI stands out for CMOS-compatible capabilities and hosting wealthy quantum registers. Our work demonstrates a critical step towards a SiC-based quantum photonic chip. We have successfully demonstrated single divacancy spin generation and nanoscale positioning on SiCOI. Using dynamical optical pumping near GSLAC, single nuclear spin polarization with a high degree of $98\%$ has been experimentally demonstrated at ambient conditions. The coherent control of a single nuclear spin and an entangled state with a fidelity of 0.89 are achieved on the SiCOI as well. Based on the nanoscale positioning techniques, we have further integrated the entangled quantum register into SiC photonic waveguides for the first time and confirmed that all the optical and spin quantum characteristics are well preserved.

Our study contributes insights into developing on-chip quantum applications using SiC color centers. The on-chip integration approach may have advantages over bulks in both of quantum sensing and quantum networks. For quantum sensing,  on-chip routing in SiC photonic structures embeds quantum sensors, eliminating the need for top laser exposure. Additionally, multi-node sensing\cite{cai2021parallel} and enhanced sensitivity\cite{castelletto2022silicon} can be achieved via photonic links and confinement. Nuclear spins in SiCOI also provide valuable sensors for nano-tesla magnetic sensing\cite{simin2016all}, atomic nuclear spin imaging\cite{ajoy2015atomic} and nuclear spin gyroscopes\cite{jarmola2021demonstration,soshenko2021nuclear}. In the case of quantum networks, photonic integrated circuits also play a pivotal role. It offers photonic links for entangling remote spins via heralded protocols\cite{bernien2013heralded,nemoto2014photonic} and enhances light-matter interactions through various photonic crystal cavities\cite{riedel2017deterministic,bracher2017selective,lukin2023two,crook2020purcell,soref2022classical}, vital to increasing the rate of entanglement generation.  We believe that our work shall shed light on spin-photon entanglement generation\cite{gao2012observation}, remote entanglement of two or more SiC spins, and the eventual realization of multi-node quantum networks on a chip.

\section*{Methods}

\textbf{4H-SiCOI and photonic devices fabrication.}
After the standard RCA cleaning, tens of nanometers thermal SiO2 was grown on a 4-inch 4H-SiC wafer with an epitaxy layer (The N-doping of the 4H-SiC wafer and epi-layer were 1×10$^{18}$ and 5×10$^{14}$ cm$^{-3}$ respectively). Then the wafer and a oxidized Si substrate were activated with 100 W O$_{2}$ plasma for 30 s. Then, the two wafers were bonded under the pressure of 3000 N at room temperature. Before the grinding process, the bonded structure was further annealed at 800 $^{\circ}$C for 6 hours to enhance the bonding strength. The grinding process was divided into two steps. Firstly, the bonded wafer was treated with 10 $\mu$m diamond slurry under the pressure of 600 N for about 10 hours. The SiC layer was removed to about 30 $\mu$m after that. Then, the SiC layer was further ground to a 10 $\mu$m thickness with 3 $\mu$m diamond slurry. The CMP process was necessary to remove the damaged layer due to the grinding process. Finally, the SiCOI wafer was cut into dies and treated with ICP dry etching one by one for the final thickness adjustment. Etching SiC was done with 100 W RF and 1000 W ICP power under 10 mTorr. To fabricate single divacancy spins, hole arrays were patterned on the SiC layer using 180 nm thick PMMA (A4 with rotation speed 4000 rpm) as a photoresist through E-beam lithography. Following the 30 keV Carbon ion implantation with a dose of 7.8$\times$10$^{11}$ cm$^{-2}$, the sample was annealed in a high vacuum at 900$^{\circ}$C for 30 min to remove residual lattice damage. 

Photonic device fabrication began with the deposition of 30 nm Cr on SiCOI containing single spins surrounded by alignment cross markers (5 nm Ti and 50 nm Au). Then, the devices are patterned into an electron beam resist (ZEP 520A) via an electron beam lithography overlay process. After resisting development, the Cr mask is formed by ICP dry etching. The photonic devices are obtained under the Cr hard mask by ICP etching. Etching SiC was done with 100 W RF and 1000 W ICP power under 10 mTorr.

\textbf{Optical characterization and spin manipulation.}
Single defects were characterized by a home-built confocal microscope operating at room temperature. A 914 nm CW excitation laser within the range of optimal excitation wavelengths for VV$^{0}$s \cite{wolfowicz2017optical} was focused on the single defect through a Nikon high numerical aperture oil objective (Model No.: CFI Plan Fluor 100X Oil). The fluorescence was finally counted by a superconducting nanowire single photon detector (SNSPD, Photon Technology) after passing a dichroic mirror (Semrock, Di02-R980-25×36) and a 1000 nm long-pass filter(Thorlabs, FELH1000). The SNSPD was replaced with an InGaAs CCD camera in the direct imaging system. All schematic diagrams of the setups are displayed in Supplementary Note 1. For the ODMR, Rabi, and Ramsey measurements, the microwave was generated using a synthesized signal generator (Mini-Circuits, SSG-6000 RC) and then gated by a switch (Mini-Circuits, ZASWA-2-50DR+). After amplification (Mini-Circuits, ZHL-25W-272+), the microwave signals were fed to a 20-$\mu$m-wide copper wire above the sample's surface. The excitation laser was modulated using an acoustic-optic modulator. The timing sequence of the electrical signals for manipulating and synchronizing the laser, microwave, and counter was generated using a pulse generator (SpinCore, PBESRPRO500). The external magnetic field was applied with a permanent magnet along the 4H-SiC c axis. Otherwise, the large transverse magnetic field may obscure the effect of the off-diagonal hyperfine elements. The strength of the magnetic field was adjusted by the distance of the magnet from the laser focus point on the sample. The magnetic field was determined precisely using ODMR spectrums from a single PL6 electron spin as a magnetometer.

\textbf{Single-defect imaging and nanoscale positioning.}
For positioning the PL6A single defect, a direct imaging system was used to acquire the defect array of single spins with the fabricated cross markers. Subsequently, the central highlight spot from unfiltered noise was removed by sigma clipping. After that, the processed image was cross-correlated with the ideal cross-mark pattern, and the peaks from the cross-correlation were extracted using Point Spread Function (PSF) fitting, enabling us to acquire the transformed coordinates of the corresponding cross marks. We determined the coordinates of the single defect by employing Maximum Likelihood Estimation (MLE). This allowed us to ascertain the relative positional relationship between the single defect and the cross marks (Supplementary Note 1 and Supplementary Fig. 3 for more details).

\textbf{Probability of electron-nuclear spin flip-flop.}
The nuclear polarization P is related to the probability of electron-nuclear spin flip-flop, which can be calculated by the degree of mixing between spin and energy eigenstates. In our work, the probability of nuclear spin up (${\rho }_{\uparrow }$) and down (${\rho }_{\downarrow }$) flips are calculated by relevant eigenstates\cite{wang2013sensitive}:
\begin{gather}
{{\rho }_{\uparrow }}=4\cdot {{\left| \left\langle 0,\downarrow  \right.\left| \beta  \right\rangle  \right|}^{2}}\cdot \left[ {{\left| \left\langle 0,\downarrow  \right.\left| \chi  \right\rangle  \right|}^{2}}+{{\left| \left\langle 0,\downarrow  \right.\left| \delta  \right\rangle  \right|}^{2}} \right]+4\cdot {{\left| \left\langle 0,\downarrow  \right.\left| \delta  \right\rangle  \right|}^{2}}\cdot \left[ {{\left| \left\langle 0,\downarrow  \right.\left| \alpha  \right\rangle  \right|}^{2}}+{{\left| \left\langle 0,\downarrow  \right.\left| \chi  \right\rangle  \right|}^{2}} \right]
\label{eq:2}
\\
{{\rho }_{\downarrow }}=4\cdot {{\left| \left\langle 0,\uparrow  \right.\left| \alpha  \right\rangle  \right|}^{2}}\cdot \left[ {{\left| \left\langle 0,\uparrow  \right.\left| \chi  \right\rangle  \right|}^{2}}+{{\left| \left\langle 0,\uparrow  \right.\left| \delta  \right\rangle  \right|}^{2}} \right]+4\cdot {{\left| \left\langle 0,\uparrow  \right.\left| \chi  \right\rangle  \right|}^{2}}\cdot \left[ {{\left| \left\langle 0,\uparrow  \right.\left| \beta  \right\rangle  \right|}^{2}}+{{\left| \left\langle 0,\uparrow  \right.\left| \delta  \right\rangle  \right|}^{2}} \right]
\label{eq:3}
\end{gather}

where $\left|\alpha \right\rangle,\left|\beta \right\rangle,\left|\chi \right\rangle and \left|\delta \right\rangle$ are the eigenstates of Hamiltonian. The projections onto the ${\ket{0,\uparrow/\downarrow}}$ states of these eigenstates are affected by the off-diagonal hyperfine elements.

\textbf{Entanglement fidelity.}
To determine the entanglement fidelity, we compare the measured density matrix $\rho$ with the ideal target density matrix $\rho'$ using the following definition\cite{bourassa2020entanglement}: 

\begin{equation}
\label{eq:4}
F = Tr(\sqrt{\sqrt{\rho}\rho'\sqrt{\rho}})
\end{equation}

\section*{Data availability.}
The data that support the findings of this study are included within the paper and its Supplementary Information file. Source Data are provided with this paper. Any other relevant data are available from the corresponding authors upon request.

\section*{Code availability.}
The codes used for plotting the data are available from the corresponding authors on request.


\section*{Acknowledgments}
We thank Fazhan Shi, Jin Liu and Gangqin Liu for the fruitful discussions. We acknowledge the support from the National Key R$\&$D Program of China (Grant No. 2021YFA1400802, 2022YFA1404601, 2023YFB2806700), the National Natural Science Foundation of China
(Grant No. 12304568, 11934012, 62293520, 62293522, 62293521, 12074400 and 62205363), the GuangDong Basic and Applied Basic Research Foundation (Grant No. 2022A1515110382), Shenzhen Fundamental research project (Grant No. JCYJ20230807094408018), Guangdong Provincial Quantum Science Strategic Initiative (Grant No. GDZX2403004, GDZX2303001, GDZX2306002, GDZX2200001), Y.Z. acknowledges the support from the Young Elite Scientists Sponsorship Program by CAST, Q.S. acknowledges the support from the New Cornerstone Science Foundation through the XPLORER PRIZE, A.Y. acknowledges the support from the Shanghai Science and Technology Innovation Action Plan Program (Grant No. 22JC1403300) and CAS Project for Young Scientists in Basic Research (Grant No. YSBR-69), Z.L. acknowledges the support from the Major Key Project of PCL, the Talent Program of Guangdong Province (Grant No. 2021CX02X465).
\section*{Author contributions}

Y.Zhou and Q.S. conceived the idea. A.Y. and X.O. prepared the SiC membrane. H.H., Y.Zhang, Q.Luo, S.X., and Z.W. carried out the EBL lithography and SiC defect generation. Y.Zhou and H.H. built the setup and carried out the measurements. H.H., T.B., Y.Zhou, and Q.S. performed the simulations. Y.Zhou, H.H., Q.Li, D.L., C.L., and Z.L. contributed to the data analysis.  All authors contributed to writing the paper.

~\\

\section*{Competing interests}

The authors declare no competing interests.

\begin{figure}[ht]
\centering
\includegraphics[width=0.9\linewidth]{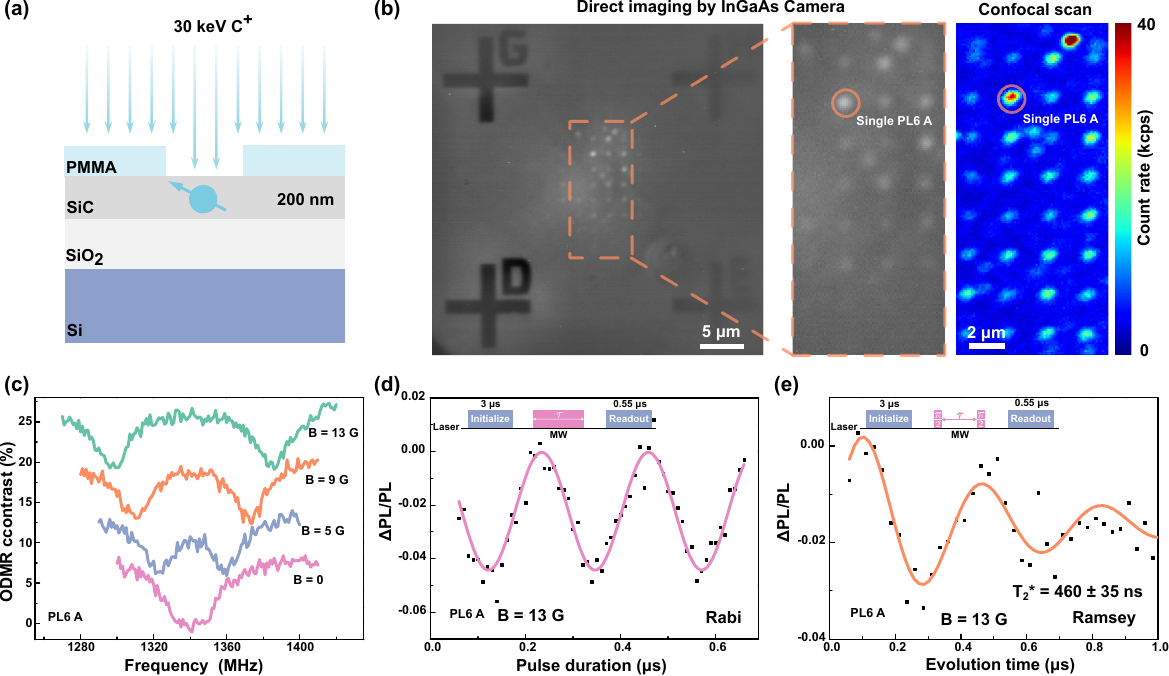}
\caption{Generation and coherent control of single color center PL6 A on SiCOI. (a) Diagram of the cross-section view of the SiCOI sample with 30 keV $C^{+}$ ion implantation through a PMMA mask with an 8$\times$8 array of 100 nm diameter holes. (b) Direct and confocal scanning image of the implanted sample. The Au cross marker is deposited for nanoscale spatial positioning. PL6 A is circled in both the CCD image and the confocal scan. (c) ODMR spectra of PL6 A under varying c-axis magnetic fields, with each ODMR peak displaying a divergence at a slope of 2.82 $\pm$ 0.02 $\rm{MHz/G}$. D = 1340.4 $\pm$ 1.3 MHz, E = 6.95 $\pm$ 1.82 MHz are extracted from the peaks' fitting.(d) Rabi oscillation and (e) Ramsey fringes of the PL6 A under 0.1 mW excitation. Each cycle is repeated around 10 million times to reduce the measurement error. The black dots represent the raw data, and the solid lines depict the fitting with the cosine and cosine exponential decay functions, respectively. From the fitting, the inhomogeneous spin-dephasing time T$^{*}_{2}$ is extracted to be $460 \pm 35 $ ns. }
\label{fig:stream}
\end{figure}

\begin{figure}[ht]
\centering
\includegraphics[width=0.9\linewidth]{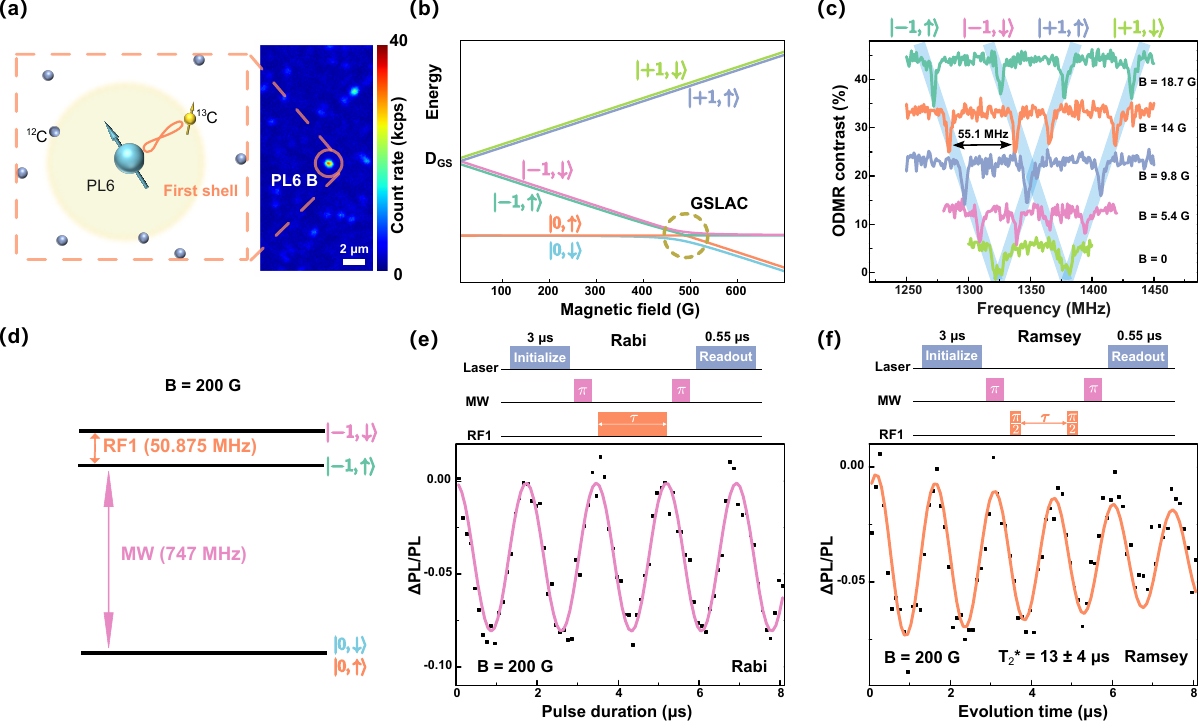}
\caption{Coherent control of single  $^{13}$C nuclear spins coupled with PL6 B on SiCOI. (a) An illustration of a single PL6 electron spin coupled with a $^{13}$C nuclear spin in the first shell. (b) The energy diagram of ground state spin sublevels as a function of c-axis magnetic field B. GSLAC is marked with a dashed circle.We use 0, ±1, and $\uparrow$, $\downarrow$ to denote the electron and nuclear spin, respectively.(c) ODMR spectra of the PL6 B electron spin coupled with the first shell $^{13}$C nuclear spin from 0 G to 18.7 G. The pale blue lines indicate the change of different peaks with the magnetic field. The slop is  2.82 $\pm$ 0.03 $\rm{MHz/G}$, corresponding to a c-axis defect's Zemman splitting. (d) Simplified representation of the energy Levels of the electron-nuclear coupled system at B = 200 G. ${\ket{0,\uparrow}}$ and ${\ket{0,\downarrow}}$ are degenerate. (e) Rabi and (f) Ramsey of the $^{13}$C nuclear spin under 0.1 mW excitation. Their pulse sequences contain laser initialize, MW $\pi$-pulse, and RF1 $\pi$-pulse or $\pi/2$-pulse. MW and RF1 are the pulses that drive ${\ket{0,\uparrow\downarrow}}$ to ${\ket{-1,\uparrow}}$ and ${\ket{-1,\uparrow}}$ to ${\ket{-1,\downarrow}}$ respectively. Each cycle is repeated around 10 million times to reduce the measurement error. The nuclear spin is mapped and read out through the electron spin. The Ramsey fringe is fitted with a cosine exponential decay function, yielding a T$_{2}^{\ast}$ = 13 $\pm$ 4 $\mu$s.}
\label{fig:stream}
\end{figure}

\begin{figure}[ht]
\centering
\includegraphics[width=0.55\linewidth]{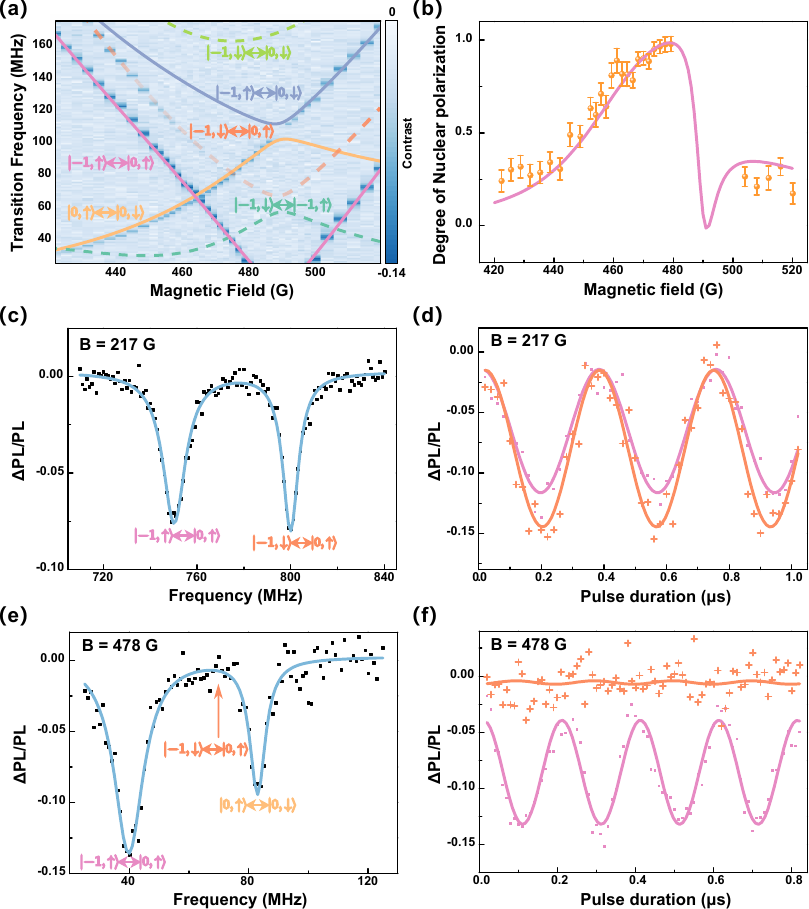}
\caption{Dynamic polarization of the single  $^{13}$C nuclear spin (PL6 B) on SiCOI. (a) ODMR spectra of the single PL6 defect coupled with single first-shell $^{13}$C nuclear spin at different magnetic fields near GSLAC. The vibrant lines are the theoretical calculation of the transition frequency of the spin sublevels with the total Hamiltonian \ref{eq:1}.${\ket{-1,\uparrow}}$ to ${\ket{0,\uparrow}}$ maintains its intensity as the magnetic field approaches the anticrossing point. In contrast, ${\ket{-1,\downarrow}}$ to ${\ket{0,\uparrow}}$ weakens and vanishes from B = 460 G, signifying where strong nuclear polarization occurs. (b) The degree of nuclear spin polarization is defined by $P=({I^+}-{I^-})/({I^+} +{I^-})$, and ${I^+}$ and ${I^-}$ are extracted from the Lorentzian fitting of the ODMR spectra in (a). The maximum polarization degree is $0.98\pm 0.04$. The error bars are propagated from the standard deviations of parameters in the Lorentzian fitting. (c) and (e)  ODMR spectra recorded for B = 217 G (away from GSLAC and no obvious nuclear polarization) and B = 478 G (at GSLAC where nuclear polarization is strong). The peaks correspond to transitions $\left|-1\right.,\left.\uparrow\right\rangle\leftrightarrow\left|0\right.,\left.\uparrow\right\rangle$ and $\left|-1\right.,\left.\downarrow \right\rangle\leftrightarrow\left|0\right.,\left.\uparrow\right\rangle$ respectively. (d)$\&$(f) Rabi oscillations of the transitions in (c)$\&$(e).}
\label{fig:stream}
\end{figure}

\begin{figure}[ht]
\centering
\includegraphics[width=0.9\linewidth]{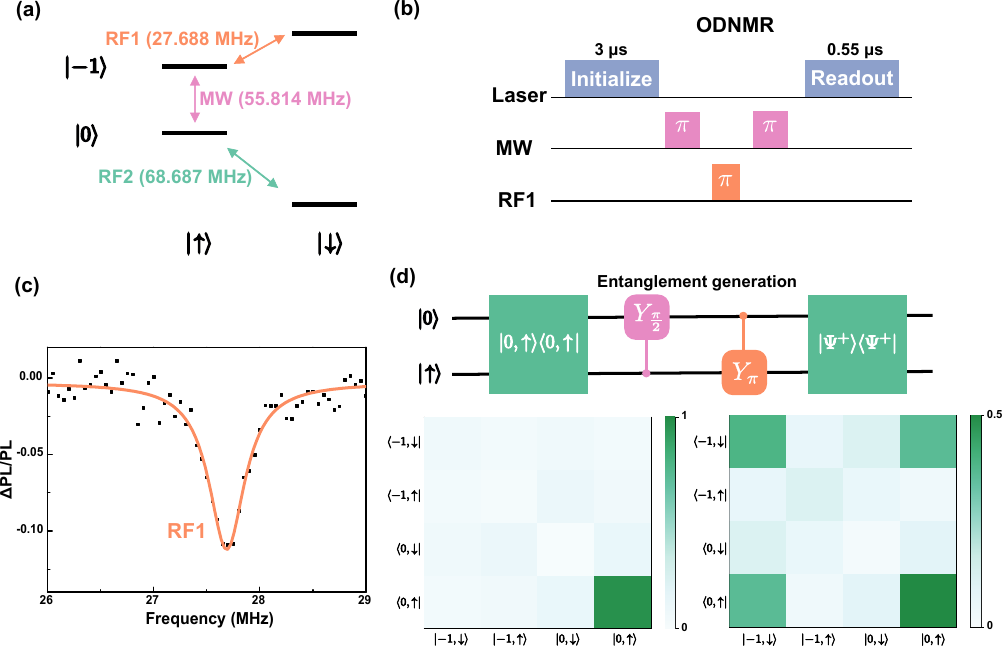}
\caption{Optically detected nuclear magnetic resonance (ODNMR) and entanglement generation of PL6 B. (a) Simplified energy level scheme of the electron-nuclear coupled system. MW, RF1, and RF2 are three addressable transitions. MW and RF2 can be obtained directly from the ODMR spectra, while RF1 is missing due to the strong nuclear polarization. To recover RF, the pulse sequence in (b) is used with the help of electron spin. (c) Optically detected nuclear magnetic resonance measurement with the sequence in (b). (d) upper panel: Gate operation circuit diagram for the entanglement generation and state tomography. lower panel: density matrix from quantum state tomography of the initial ($|\left.0,\uparrow\right\rangle\left\langle0,\uparrow\right.|$) and entangled state ($|\left.\Psi^+\right\rangle\left\langle\Psi^+\right.|$ ). The fidelity of the target state is 0.98 and 0.89, respectively.}
\end{figure}

\begin{figure}[h]
\centering
\includegraphics[width=0.85\linewidth]{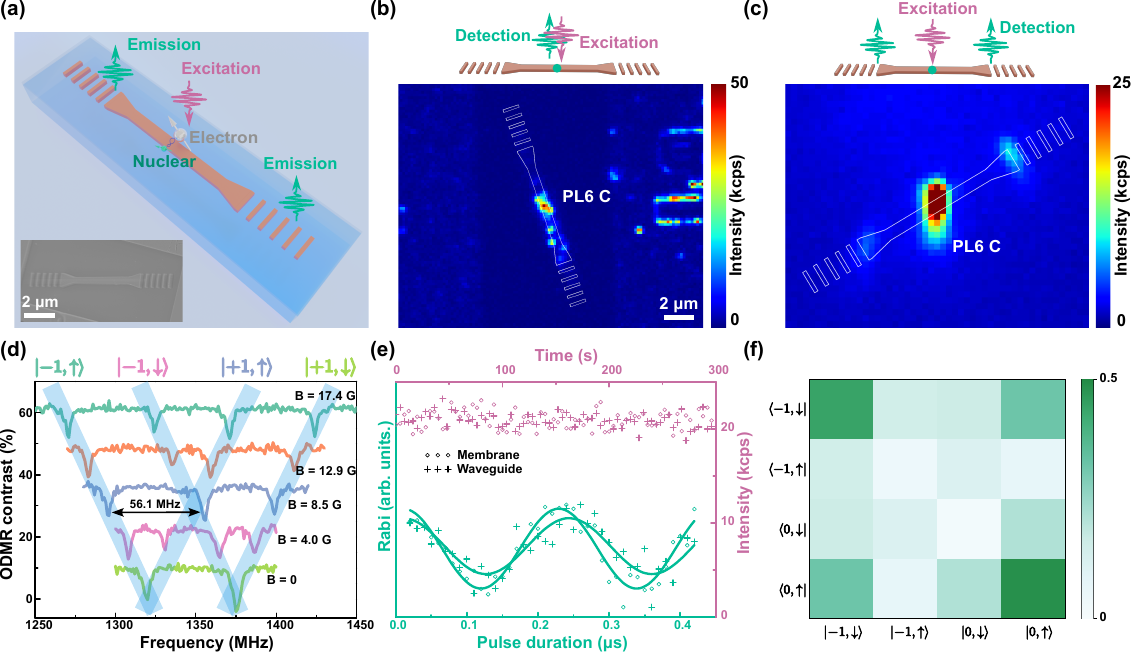}
\caption{Deterministic integration of the electron-nuclear quantum register PL6 C into waveguides. (a) Schematic of the waveguide with two directional couplers, showing the quantum register at the center. Inset: Scanning Electron Microscope (SEM) image of the fabricated waveguides. (b) Confocal scan map of PL6 C after integration, with the outline of the waveguide included to guide the eye. (c) Collection angle-dependent scan map while PL6 C is excited. The central spot aligns with the defect, while the adjacent two spots are emissions from the directional couplers. The outline of the waveguide is included to guide the eye. (d) ODMR spectra of the waveguide-integrated PL6 C electron spin coupled with the first shell $^{13}$C nuclear spin in different magnetic B fields. The pale blue lines indicate the change of different peaks with the magnetic field. The slope is 2.79 $\pm$ 0.01 $\rm{MHz/G}$, corresponding to the Zeeman splitting of a c-axis defect. The four transitions correspond to ${\ket{-1,\uparrow}}$, ${\ket{-1,\downarrow}}$, ${\ket{1,\uparrow}}$, and ${\ket{1,\downarrow}}$, respectively. The coupling strength is 56.1 $\pm$ 0.2 MHz. (e) Comparison of photon stability (0.2 mW CW excitation, with an interval of 0.1 s and duration of 300 s) shown by purple hollow rhombuses (before integration) and crosses (after integration). Rabi oscillations (MW transition) are indicated by green hollow rhombus (before integration) and cross (after integration). (f) The density matrix of the generated Bell state with the quantum registers in a waveguide, showing a fidelity of 0.88 from quantum state tomography.}
\end{figure}

\end{document}